\documentclass[aps,prd,twocolumn,superscriptaddress,showpacs]{revtex4}
\usepackage{graphicx}
\usepackage{dcolumn}
\usepackage{bm}
\usepackage{natbib}

\newcommand{\apjs}{{Astrophys.~J.~Supp.}}

\newcommand{\be}{\begin{equation}}
\newcommand{\ee}{\end{equation}}
\newcommand{\bea}{\begin{eqnarray}}
\newcommand{\eea}{\end{eqnarray}}

\begin{document}
\title{An indirect limit on the amplitude of primordial Gravitational
  Wave Background from CMB-Galaxy Cross Correlation}

\author{Asantha Cooray}
\email{acooray@uci.edu}

\affiliation{Dept. of Physics \& Astronomy
 University of California,
 Irvine, CA 92697-4575}

\author{Pier-Stefano Corasaniti}
\email{pierste@phys.columbia.edu}

\affiliation{ISCAP, Columbia University,
550 West 120th Street,
New York, NY, 10027 (USA)}

\author{Tommaso Giannantonio}
\email{tommaso.giannantonio@roma1.infn.it}
\affiliation{Dipartimento di Fisica ``G. Marconi'', Universita'
di Roma ``La Sapienza'', Ple Aldo Moro 5, 00185, Rome, Italy.}

\author{Alessandro Melchiorri}
\email{alessandro.melchiorri@roma1.infn.it}
\affiliation{Dipartimento di Fisica ``G. Marconi'' and INFN, sezione
  di Roma, Universita' di Roma ``La Sapienza'', Ple Aldo Moro 5,
  00185, Roma, Italy.}

\begin{abstract}
While large scale cosmic microwave background (CMB) anisotropies involve a combination of the
scalar and tensor fluctuations, the scalar amplitude can be independently determined through
the CMB-galaxy cross-correlation. Using recently measured cross-correlation amplitudes,
arising from the cross-correlation between galaxies and the 
Integrated Sachs Wolfe effect in CMB anisotropies, we obtain a constraint
 $r < 0.5$ at 68 \% confidence level 
on the tensor-to-scalar fluctuation amplitude ratio. 
The data also allow us to exclude gravity waves at a level of a few percent, relative to the
density field, in a low - Lambda dominated universe ($\Omega_{\Lambda} \sim 0.5$). 
In future, joining cross-correlation ISW measurements, which captures cosmological parameter information, with 
independent determinations of the matter density and CMB anisotropy power spectrum, may
constrain the tensor-to-scalar ratio to a level above 0.05. This value is the ultimate limit on tensor-to-scalar ratio 
from temperature anisotropy maps when all other cosmological parameters except for the
tensor amplitude are known and the combination with CMB-galaxy correlation allows this limit to be
reached easily by accounting for degeneracies in certain cosmological parameters.

\end{abstract}
\maketitle

\section{Introduction}
\label{sec:intro}

One of the major predictions of inflation is the existence of a 
stochastic background of gravitational waves (GW)
(see e.g. \cite{turner}, \cite{dodelson}).
The amplitude of these tensor fluctuations is
proportional to the square of the energy scale 
of inflation (see e.g. \cite{krauss}). 
Furthermore, the `tilt'
of the GW spectrum (and of the scalar as well) 
can give us direct information on the slope of
the inflaton potential up to second derivatives.
Therefore the detection of this stochastic background would 
provide valuable information on the inflationary scenario
sheding light on the physics at $\sim 10^{16}$ GeV (see e.g. \cite{hoffman}).

The stochastic GW background leaves a signature on the CMB
through different mechanisms. In principle a clear detection can be 
obtained from measurements of the B-polarization. 
Tensor perturbations also induce temperature fluctuations 
through a Sachs-Wolfe effect and an integrated one 
(for a review see \cite{Pritchard}). 
The former is localized by the visibility function
at the last scattering surface and consequently 
depends on the recombination history. 
Since it occurs in a relatively short
period of time the amplitude of this effect is negligible at all scales. 
On the contrary GW produce most of the signal through the integrated effect
which is sensitive to the evolution
of gravitational waves from the time of last scattering to today.
Since the observed temperature anisotropy power spectrum is the sum of both
the scalar and tensor components, 
for relatively low amplitudes of the tensor modes 
the scalar fluctuations dominate the spectrum and 
the tensor contribution is dwarfed.
Similarly, if there is a non-negligible contribution from GWs,
the sum of scalar and tensor spectra would lower the predicted
amplitude of the acoustic peaks
relative to the angular scales of few or more degrees. 
Thus, the ratio of tens of degree to sub-degree fluctuations 
can be used to constrain the presence of GW.

With the advent of WMAP, 
several studies have addressed whether there is a significant contribution of
GWs in the CMB spectra (see e.g. \cite{peiris,kinney,barger,leach}). 
Despite the high quality of these data, it still remains difficult to obtain
a robust constrain on the amplitude of the tensor perturbations 
due to the degeneracies with other cosmological parameters. 
For instance, both the scalar spectral index 
$n_S$ and its running are strongly degenerate with the
amplitude of the GW background. In \cite{Spergel}, combining the 
CMB data with the 2dF matter power spectrum yield a $2 \sigma$ constraint
on the $tensor-to-scalar$ ratio $r<0.9$ \footnote{As in \cite{peiris}
  we define $r$ as the inflationary scalar/tensor ratio at $k_0=0.002 Mpc^{-1}$}.
The combination of WMAP data with Sloan Ly-$\alpha$ forest,
lensing bias, and galaxy clustering data constrain $r < 0.36$ at the 95\% confidence level
in a flat cosmological model \cite{seljak}.
Measurements of B-mode CMB polarization from future
experiments such as the {\it Planck} satellite, 
or in the long-term the CMBpol mission
as part of NASA's {\it Beyond Einstein} Program, will provide either a 
detection of the tensor modes or the best 
constrain on their amplitude \cite{GWpolB}. Concept studies are already under
way to consider the possibility for a direct detection of the relic gravitational wave
background present today with space-based Laser interferometers as part of a second-generation mission to
followup NASA's Laser Interferometer Space Antenna  (LISA; \cite{tristan}).
In the meantime, it is worthwile to investigate methods that could 
independently bound the GW contribution and/or break some 
of the degeneracies which affect present constraints.
This would also be useful in planning observational
strategies of current and upcoming CMB experiments.

The scalar and tensor temperature anisotropy power  spectra are characterized by a nearly
constant plateau on the large angular scales. The temperature anisotropy power specrum are
sums of tensor and scale contributions, $C_l^{TT} = C_l^{TT,scalar}+C_l^{TT,tensor}$.
To establish the amplitude of tensor fluctuations from total temperature anisotropies,
which is what we observe, one must need to properly account for the contribution from
scalar fluctuations.  Thus,  an independent estimate of the scalar amplitude, when
combined with temperature anisotropy spectrum, can be used to constrain  the amplitude
of tensor fluctuations, or, the ratio of tensor-to-scalar fluctuation amplitude.
We should always keep in mind that due to the integrated Sachs-Wolfe effect produced at late time 
by the scalar perturbations, $C_l^{TT,scalar}$ has a non-trivial normalization that depends
both on the potential at last scattering (the Sachs-Wolfe effect), as well as the time evolution of the
potential; The normalization of the CMB temperature power spectrum also depends on the 
amount as well as the nature of the dark enegy given that potential fluctuations
evolve differently depending on late-time cosmology \cite{Coras,Kunz,Cooray}. 

The scalar amplitude can be inferred from the
matter power spectrum as measured by galaxy surveys \cite{tegmark,seljak2}. 
Alternatively we can make use of the positive
cross-correlation between WMAP anisotropy maps and LSS surveys 
\cite{Gaztanaga03,Boughn0304,Scranton,Nolta04,Afshordi}.
Here we consider such a possibility and determine
an independent constrain on the tensor-to-scalar ratio. 
Our limit improves those inferred from the combination of the WMAP 
data with either 2dF or SDSS surveys \cite{Spergel,tegmark}, 
by nearly a factor of two.
In particular we find that in a flat cosmological constant 
dominated universe $r<0.59$ at $2\sigma$.
We also discuss how the constraints on $r$ change by
including more accurate information on the ISW effect. 

\section{Limit on Tensor-to-Scalar Ratio}

The correlation between the large angular scale CMB temperature anisotropy
maps and the distribution of the large scale structure (LSS) is 
consequence of the integrated Sachs-Wolfe. As the universe
starts deviating from the matter dominated expansion, 
the gravitational potentials associated with clumps of
matter change in time. CMB photons crossing these regions undergo a shift
which causes a temperature anisotropy. Since the clustered structures
arises from the scalar fluctuations the CMB-LSS correlation depends
only on the scalar modes and does not get any contribution from the temperature 
anisotropies induced by the gravitational
waves background. Indeed the latter is uncorrelated with the distribution of
galaxies, cluster and super-cluster. Thus, the cross-correlation between
CMB maps and surveys can be used to infer $A_s$, under the assumption
of a given dark energy model.

The angular cross-correlation in Legendre series is given by
\be
C^X(\theta)=\sum_{l=2}^{\infty}\frac{2l+1}{4\pi}C_l^{X}P_l(\cos(\theta),
\label{cxpl}
\ee
where $P_l(\cos\theta)$ are the Legendre polynomials and $C_l^{X}$ 
is the cross-correlation power spectrum \cite{Cooray2,GarrLevon}:
\be
C_l^X=\frac{2}{\pi}\int k^2 dk\; P_{\delta \delta}(k)
I^{ISW}_l(k)I^{LSS}_l(k) \, .
\ee 
Here, $P_{\delta \delta}(k)$ is the power spectrum of density fluctuations (which
captures the amplitude of scalar fluctuations) and
integrand functions $I^{ISW}_l(k)$ and $I^{LSS}_l(k)$ are defined  as
\begin{eqnarray}
I^{ISW}_l(k)&=&\int W^{\rm ISW}(k,r) j_l[k r] dr\\
I^{LSS}_l(k)&=&\int W^{\rm LSS}(k,r) j_l[k r] dr
\label{growth}
\end{eqnarray}
when the two window functions are
\begin{eqnarray}
W^{\rm ISW}(k,r) &=& -3 \Omega_m \left(\frac{H_0}{k}\right)^2 \frac{d}{dr}\left(\frac{G}{a}\right) \nonumber\\
W^{\rm LSS}(k,r) &=&  b(k,r) n(r) G(r) \,  .
\end{eqnarray}
In above, $j_l[k r]$ is the spherical Bessel function when $r(z)$ is the
comoving distance, $b(k,r)$ is the scale- and redshift, or distance-dependent bias 
factor of galaxies with a normalized distribution, in distance, of $n(r)$ ($\int dr n(r)=1$),
and $G(r)$ is the growth function of density fluctuations. Other parameters have standard definitions.
In converting time evolving potential fluctuations, captured by the ISW effect,
to density fluctions, we have assume a flat-comsological model. If not, the window function related to
the ISW term, $W^{\rm ISW}(k,r)$, will contain corrections related to curvature.
We refer the reader to  Ref.~\cite{Cooray2,GarrLevon,Levon,Coras2,Afshordi2}
for a more detailed discussion on the cross-correlation power spectrum.

From Eq.~(\ref{growth}), it is evident that the determination of $A_s$
from cross-correlation measurements depends on the galaxy bias $b(k,z)$
of the LSS catalog used. Nevertheless, we can state in a conservative
way that the r.m.s. value of the matter fluctuation at $R=8 {\rm Mpc}/h$
lies in the range $\sigma_8=0.7-1.1$ \cite{Kunz}. Normalizing the
predicted matter-matter power spectra to these values and comparing 
to the observed galaxy-galaxy correlation function one can determine the 
corresponding bias. Marginalizing over $\sigma_8$ introduces, 
for a single catalog, an extra $\sim 20 \%$ error 
in the measurement of the ISW-correlation signal. 
This error can be further reduced if one considers several ISW
detections with catalogs affected by different biases,
or, in the case of a single survey, if biasing
can be determined as a function of the galaxy luminosity \cite{zehavi}.
Here is worth remarking that the dependence of the cross-correlation
on the bias is {\it linear}. Therefore, the error on the inference of $A_s$ 
induced by the uncertainty on the bias is fractionally less here than in the standard 
method involving the combination of CMB anisotropy data and LSS clustering spectrum (e.g. \cite{Spergel}), as
the full galaxy-galaxy correlation function 
{\it quadratically} depends on the galaxy bias factor.

Another caveat concerns the dark energy. 
Current measurements of the cross-correlation 
lack the accuracy to strongly constrain any dark energy parameters 
\cite{Coras2}. The next generation of large scale structure surveys will 
provide better constraints by measuring the correlation between the CMB 
and different redshift sample of galaxies \cite{Levon2}. As shown in
\cite{Levon,Coras2} the redshift behavior of the cross-correlation 
signal depends 
on the specific of the dark energy, while $A_s$ is an overall normalization 
factor independent of the redshift.

The cosmological constant is the simplest dark energy model 
to best fit current data \cite{Spergel,tegmark,Coras3,seljak2,riess}.

\begin{figure}[!t]
\includegraphics[scale=0.4]{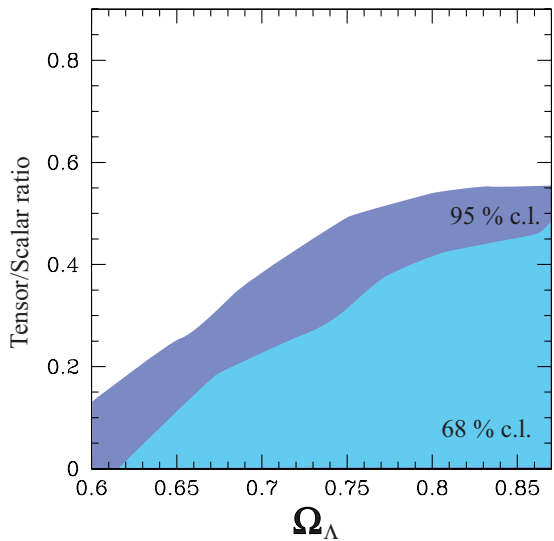}
\caption{Likelihood contours in the $\Omega_{\Lambda}-r$ plane from
current ISW detections (see \cite{gazt}). Flatness is assumed.
The ISW signal is smaller for small values of the Cosmological Constant.
}
\label{iswgw}
\end{figure}

We thefore limits our analysis to flat LCDM cosmologies and use
the current ISW detections as collected by \cite{gazt}. These data
consist of average angular cross-correlation measurements between
WMAP temperature maps and 5 different LSS surveys.
For a more detailed discussion on the evaluation of the likelihood statistics,
the effect of systematics and the correlations between redshift bins 
we refer to~\cite{Coras2}. 
We perform a likelihood analysis for the amplitude of the tensor modes $A_T$ 
and marginalize over the following parameters: the cosmological 
constant in the range $0.6< \Omega_{\Lambda}<0.9$, the
scalar spectral index $0.8 < n_s < 1.3$ and a Hubble parameter
$0.6< h <0.9$. We also vary $\sigma_8$ between 0.7 and 1.1.
We assume a spatially-flat cosmological model, and matter density changes
 as $\Omega_\Lambda$ is varied with the condition that $\Omega_m=1-\Omega_\Lambda$.
To obtains constrains on the tensor amplitude, we only make use of
large scale, $10< \ell < 40$, WMAP data. 
In order to match the region of angular scales probed
by the ISW-correlation data and avoid biasing due to
lack of power in WMAP quadrupole to the octopole, we have excluded the region
of multipole $\ell<10$. 
We also vary the baryon density, as a function of $h$, with a fixed value for $\Omega_b h^2$
of 0.02, but found little variation in our results to variations in $\Omega_b$;
This is due to the fact that the analysis excludes acoustic peaks that are more
sensitive to the exact baryon density, relative to matter, while at large scales,
corresponding to the multipole range above, there is little or no sensitivity in the
anisotropy spectrum to parameter $\Omega_b$.

We plot the results in Fig~.\ref{iswgw}, which shows the 
$1$ and $2\sigma$ contours in the $\Omega_{\Lambda}-r$ plane. 
As we may notice current data constrain the tensor 
contribution to $r <0.59$ at $2 \sigma$ confidence level. 
Since we have only $5$ weak detections of the ISW-galaxy cross-correlation,
which may be affected by unknown systematics, 
one should be careful in a direct comparison with other
constraints in the literature. 
Nonetheless the results of this analysis already illustrates the potential
of this method which captures all the information one can extract on the tensor
amplitude given the knowledge of the galaxy distribution. 
In fact, the previous standard approach combines only the CMB and matter power spectrum.
Similarly we use the galaxy power spectrum to constrain the bias parameter, but with the addition
of the CMB-LSS cross-correlation. The only way to improve this method would be
to consider the cross-correlation as a function of galaxy luminosity and redshift, 
since this would allow a better separation of the uncertainties related to bias
and dark energy respectively. 

As shown in Fig.~1, the constraint on the tensor-to-scalar
ratio depends on $\Omega_{\Lambda}$. In particular larger values of $\Omega_{\Lambda}$
allow a slightly greater tensor contribution. This trend is expected, in fact for low $\Lambda$
values the ISW-correlation is smaller, therefore to match the observed cross-correlation
larger values of the scalar amplitude are favoured and thus strongly limiting the tensor contribution
when combined with the CMB normalization. On the contrary, increasing the dark energy density
causes a larger ISW-correlation signal which will favour low values of $A_s$ and consequently
a larger contribution of the tensor modes.
If future dark energy measurements, such as from Type Ia supernovae, are to point towards values of 
the cosmological constant around $\Omega_{\Lambda} \sim 0.5 - 0.6$, 
the current cross-correlation data already constrain the GW contribution to be less than
$10 \%$ of the scalar amplitude. Beyond dark energy, we also found a minimal correlation of our constraint 
with the spectral index $n_S$ over the range of values considered. Given the limited information from 
the cross-correlation we did not vary other parameters or included
a running to the tilt. 

It is clear that accounting for the CMB-LSS cross-correlation provide better
limits on the tensor-to-scalar ratio than those derived from
WMAP data alone, from which $r<0.9$.
However the improvement does not increase significantly as one includes 
further information on the CMB-LSS cross-correlation.
To understand this point, consider the CMB temperature fluctuation at large angular scales,
\begin{equation}
\Delta T = \Delta T^{\rm SW} + \Delta T^{\rm ISW} + \Delta T^{\rm GW}.
\end{equation}
Assuming that all other cosmological parameters are known except for the amplitude
of the tensor perturbation $A_T$, the limit on the tensor-to-scalar ratio is 

\begin{equation}
 \sigma_{A_T}^{-2} = \sum_l \frac{1}{\sigma^2_{C_l}}\left(\frac{\partial C_l^{\rm GW}}{\partial A_T}\right)^2  
\end{equation}
where
\begin{equation}
\sigma_{C_l} = \sqrt{\frac{2}{2l+1}}\left(C_l^{\rm SW} + C_l^{\rm ISW} +2C_l^{\rm SW-ISW} \right) \, ,
\end{equation}
and we have assumed cosmic variance limited measurements.  Taking the full anisotropy spectrum
(which includes contributions beyond SW and ISW effects), the above estimate leads 
to a cosmic variance limit on the tensor-to-scalar ratio of 0.06 at the one-sigma confidence level. 
In order to improve this limit further, the CMB-LSS cross-correlation 
can be used to remove the ISW contribution.  The extent to which the signal improves depends on how well
$C_l^{\rm ISW}$, and to a lesser extent the correlation between SW and ISW effects $C_l^{\rm SW-ISW}$
can be removed from the data. If a fraction $c$ of the ISW effect, in tempertature maps, 
can be accounted for by the CMB-LSS cross-correlation,
then the improved noise estimate for the calculation is
\begin{equation}
\sigma'_{C_l} = \sqrt{\frac{2}{2l+1}}\left[C_l^{\rm SW} + (1-c)^2 C_l^{\rm ISW} +2(1-c)C_l^{\rm SW-ISW}\right],
\end{equation}
where we have ignored that the removal in the temperature maps also reduces the
correlated part with ISW by a similar fraction though in reality, this may be different given the projection
effects and the redshift dependence of the LSS maps used to clean the CMB maps from the ISW contribution.
Assuming that one can safely remove a 50\% contribution of the ISW effect, the improvement is at the level of 5\%. 
Therefore removing the ISW allows us to push the tensor-to-scalar limit only to the level of 0.056.
 
While future constraints are not significant, the ISW removal allows us to include information 
related to other parameters, as the ultimate limit from temperature anisotropies requires 
perfect knowledge on all cosmological parameters other than the tensor amplitude. 
Thus, there are significant gains to be made with using the CMB-LSS correlations to
improve the tensor-to-scalar limit. This is similar to the improvements one see when combining CMB data
with LSS data. Note that, in future, best limits on the tensor-to-scalar
ratio will come from CMB polarization observations, rather than temperature anisotropies.
It is well-known that
the direct signal in CMB B-models presents the best opportunity to establish the tensor contribution, 
even in the presence of a confusion with a lensing effect \cite{Kesden,Knox,Hu,CooKes,Seljak}. 

\section{Conclusions}

Large scale CMB anisotropies involve a combination of the
scalar and tensor fluctuations, we have argued that the scalar amplitude can be independently determined through
the CMB-galaxy cross-correlation, under the reasonable assumption that the large-scale clustering  of galaxies
is not affected by primordial gravitational waves. Previous limits on the tensor amplitude have considered
only the matter power spectrum to determine the scalar amplitude. We have shown that an improved
limit can be obtained by including the cross-correlation between CMB and large scale structure surveys,
in addition to the galaxy power spectrum alone.
Using recently measured cross-correlation data, we obtain a constraint
 $r < 0.5$ at 68 \% confidence level on the tensor-to-scalar ratio. 
The data also allow us to exclude gravity waves at a level of a few percent, relative to the
density field, in a low - Lambda dominated universe ($\Omega_{\Lambda} \sim 0.5$). 
This technique can account for the knowledge of additional cosmological parameters which are 
required to determined the amplitude of the tensor modes and helps converging on the ultimate 
tensor-to-scalar ratio limit of
0.05 from temperature anisotropies alone (assuming all other cosmological parameters are known).

\acknowledgments

We like to acknowledge Levon Pogosian and Enrique Gaztanaga for
discussion and help. P.S.C. is supported by Columbia Academic Quality Fund.
AM is supported by MURST through COFIN contract no. 2004027755.

\end{document}